\documentclass[twocolumn,prb,color,epsfig,superscriptaddress]{revtex4}
\usepackage[dvipdfmx]{graphicx} 

\begin{document}

\title{Pseudospin S=1 formalism and skyrmion-like excitations in the three body constrained extended Bose-Hubbard model}
\author{A.S.~Moskvin}
\affiliation{Ural Federal University, 620083 Ekaterinburg,  Russia}
\date{\today}


\begin{abstract}
We have focused in the paper on the most prominent and intensively studied S=1 pseudospin formalism for extended bosonic Hubbard model (EHBM) with truncation of the on-site Hilbert space to the three lowest occupation states n = 0, 1, 2. The EHBM Hamiltonian is a paradigmatic model for the highly topical field of ultracold gases in optical lattices. Generalized non-Heisenberg effective pseudospin Hamiltonian does provide a deep link with boson system and physically clear description of "the myriad of phases"\,  from uniform Mott insulating phases and density waves to two types of superfluids and supersolids. We argue that the 2D pseudospin system is prone to a topological phase separation and focus on several types of unconventional skyrmion-like topological structures in 2D boson systems, which have not been analysed till now. The structures are characterized by a complicated  interplay of insulating and the two superfluid phases with a single boson and two-boson condensation, respectively. 
\end{abstract}

\maketitle


\section{Introduction}

Since 1989 , the bosonic Hubbard model (see Refs.\onlinecite{Fisher} and references therein) has attracted continued interest due to its very rich ground state phase diagram and large opportunities of a direct experimental realization in systems of ultracold bosonic atoms
loaded in optical lattices. Such systems offer unique opportunities for studying strongly correlated quantum matter in a highly controllable environment.

The Hamiltonian of the extended bosonic Hubbard model (EBHM) is usually defined as follows
$$
H = -\sum_{i>j}t_{ij}	({\hat b}_i^{\dagger}{\hat b}_j+h.c.)+\frac{U}{2}\sum_{i}{\hat n}_{i}({\hat n}_{i}-1)+
$$
\begin{equation}
\sum_{i>j}V_{ij}{\hat n}_{i}{\hat n}_{j}-\mu \sum_{i}{\hat n}_{i}  \, ,
\label{BH}
\end{equation}
where ${\hat b}_i^{\dagger}$, ${\hat b}_i$, ${\hat n}_i= {\hat b}_i^{\dagger}{\hat b}_i$
are, respectively, the boson creation, annihilation, and number operators at the
lattice site $i$. The boson transfer amplitudes are given by $t_{ij}$; $U_i=U$ and $V_{ij}$ parametrize the Coulomb repulsions between bosons resting at the same and different sites. While $t_{ij}$ causes the bosons to delocalize, promoting a superfluid (SF) phase at weak interactions, $U$ and $V_{ij}$ tend to stabilize  the conventional Mott insulator (MI) and the density wave (DW) phases when the interaction dominates over the hopping energy scale set by $t$.

Attractive on-site  boson-boson interactions allow for the formation of dimers, or bound states of two bosons. The phase diagram then contains the conventional one-boson superfluid (1-BS) with nonvanishing order parameters $\langle {\hat b}_j\rangle \not=0$ and $\langle {\hat b}_j^2\rangle \not=0$ and  dimer superfluid (2-BS) phase. The 2-BS phase is characterized by the vanishing of the one-boson order
parameter ($\langle {\hat b}_j\rangle =0$) but nonzero pairing correlation ($\langle {\hat b}_j^2\rangle \not=0$). Apart from the above local order parameters, one can use superfluid stiffness to identify the superfluid states. It is worth noting, the thermal transitions between the 2-BS dimer superfluid and the 1-BS normal fluid are considered in Ref.\onlinecite{Kwai}. 

When the inter-site boson-boson repulsion is turned on, in addition to the uniform Mott
insulating state and two superfluid phases, a dimer checkerboard solid
state appears at unit filling, where boson pairs form a solid with checkerboard structure.

Our starting point for theoretical analysis of  the 2D extended Bose Hubbard model will assume truncation of the on-site Hilbert space to the three lowest occupation states n = 0, 1, 2 with further mapping of the EBHM Hamiltonian to an anisotropic spin-1 model (see, e.g., Refs.\onlinecite{Berg}). 
The simplest effective spin-1 model Hamiltonian is
$$
	{\hat H}=-\sum_{i>j}t_{ij}(S_{ix}S_{jx}+S_{iy}S_{jy})+\frac{U}{2}\sum_iS_{iz}^2 +
$$	
	\begin{equation}
	\sum_{i>j}V_{ij} S_{iz}S_{jz}- \mu\sum_iS_{iz}\, .
	\label{Hs}
\end{equation}

In this space the DW phase corresponds to an antiferromagnetic ordering of the pseudospins in the $z$ direction. The MI ground state, on the other hand, includes
a large amplitude of the state with $M_S=0$ on every site
with small admixture of states containing tightly bound
particle-hole fluctuations ($M_S=\pm 1$ on nearby sites). The phase can be termed as a quantum paramagnet. The 1-BS and 2-BS superfluid phases correspond to dipole and quadrupole (nematic) pseudospin XY-order, respectively. Generally speaking, one may anticipate the emergence of so-called supersolid phases, or mixed 1-BS+DW (2-BS+DW) phases.

In this paper, we do consider the most general form of the effective S=1 pseudospin Hamiltonian related with the extended Bose-Hubbard model and present a short overview of different phase states. We focus on several types of unconventional skyrmion-like topological structures in 2D boson systems, which have not been analysed till now. The structures are characterized by a complicated  interplay of insulating and two superfluid phases. The rest of the paper is organized as follows. Sec.II is devoted to introduction into the pseudospin formalism, in Sec.III we introduce and analyse the effective pseudospin Hamiltonian. In Sec.IV we turn to a short overview of a typical simplified S=1 spin model. Unconventional pseudospin topological structures are considered in Sec.V with a short conclusion in Sec.VI.

\section{Pseudospin formalism}

One strategy to handle with the physics of the extended Bose-Hubbard model with the truncated on-site Hilbert space $n=0,1,2$ is to make use of a S=1 pseudospin formalism\,\cite{MM,Batista+Ortiz} and to create model pseudospin Hamiltonian that can reasonably well reproduce both the ground state and important low-energy excitations of the full problem. Standard pseudospin formalism represents a variant of the equivalent operators technique widely known in different physical problems from classical and quantum lattice gases, binary alloys, (anti)ferroelectrics,.. to neural networks. The formalism starts with a finite basis set for a lattice site (triplet in our model). Such an approach differs from well-known pseudospin-particle transformations akin Jordan-Wigner\,\cite{JW} or Holstein-Primakoff\,\cite{HP} transformation which establish a strict linkage between pseudospin operators and the creation/annihilation operators of the Fermi or Bose type. The pseudospin formalism generally proceeds with a truncated basis and does not imply a strict relation to boson operators that obey the bosonic commutation rules.
  
The three on-site Fock states $\left|n=0\right\rangle,\left|n=1\right\rangle,\left|n=2\right\rangle$ can be addressed to form a local Hilbert space of the semi-hard core bosons which 
can be mapped onto a system of S = 1 centers via a generalization of the Matsubara-Matsuda transformation\,\cite{Batista+Ortiz} that also maps the boson density into  the local magnetization: $n_j = S_{zj} + 1$. In contrast to the hard-core bosons associated with S = 1/2
magnets, it is possible to study "Hubbard-like"\, bosonic gases with on-site density-density (contact) interactions because $n_j \leq 2$. Hereafter, we associate the three on-site Fock states with the occupation numbers $n=0,1,2$
  with the three components of the $S=1$ pseudo-spin (isospin)
triplet with  $M_S =-1,\,0,\,+1$, respectively.
It is worth noting that very similar S=1 pseudospin formalism was suggested recently\,\cite{Moskvin-11,JPCM} to describe the triplet of Cu$^{1+}$, Cu$^{2+}$, Cu$^{3+}$ valence states in copper high-temperature superconductors.

The $S=1$ spin algebra includes the three independent irreducible tensors
${\hat V}^{k}_{q}$ of rank $k=0,1,2$ with one, three, and five components,
respectively, obeying the Wigner-Eckart theorem \cite{Varshalovich}
\begin{equation}
\langle SM_{S}| {\hat V}^{k}_{q}| SM_{S}^{'}\rangle=(-1)^{S-M_{S}} \left(
\begin{array}{ccc}S&k&S
\\
-M_{S}&q&M_{S}^{'}\end{array}\right)\left \langle S\right\| {\hat
V}^{k}\left\|S\right\rangle. \label{matelem}
\end{equation}
Here we make use of standard symbols for the Wigner coefficients and reduced
matrix elements. In a more conventional Cartesian scheme a complete set of the
non-trivial pseudo-spin operators would include both ${\bf   S}$ and a number
of symmetrized bilinear forms $\{S_{i}S_{j}\}=(S_{i}S_{j}+S_{j}S_{i})$, or
spin-quadrupole operators, which are linearly coupled to $V^{1}_{q}$ and $V^
{2}_{q}$, respectively
$$
V^{1}_{q}=S_{q}; S_{0}=S_{z}, S_{\pm}=\mp \frac{1}{\sqrt{2}}(S_{x}\pm iS_{y} ):
$$
\begin{equation}
V^{2}_{0} \propto (3S_{z}^{2}-{\bf  S}^2), V^{2}_{\pm 1}\propto (S_z S_{\pm}+
S_{\pm}S_z), V^{2}_{\pm 2}\propto S_{\pm}^2 .
\end{equation}

Instead of the three $|1M\rangle $ states one may use the Cartesian basis set ${\bf \Psi}$, or $|x,y,z\rangle$:
\begin{equation}
	|10\rangle = |z\rangle\,,|1\pm 1\rangle = \mp\frac{1}{\sqrt{2}}(|x\rangle \pm i|y\rangle ) 
\end{equation}
so that an on-site wave function can be written in the matrix form as follows\,\cite{Nadya}:
\begin{equation}
	\psi=\pmatrix{c_1\cr c_2\cr
	c_3}=\pmatrix{R_1\exp(i\Phi_1)\cr R_2\exp(i\Phi_2)\cr
	R_3\exp(i\Phi_3)}\,;\qquad |\vec R|^2=1\, , \label{fun}
\end{equation}
 with ${\bf R}=\{\sin\Theta\cos\eta ,\sin\Theta\sin\eta ,\cos\Theta\}$. Obviously, the minimal number of dynamic variables describing an isolated on-site $S=1$ (pseudo)spin center
  equals to four, however, for a more general situation, when
  the (pseudo)spin system represents only the part of the bigger system, and we are
  forced to consider the coupling with the additional degrees of freedom,  one should consider all the five non-trivial parameters.

The pseudospin matrix  has   a very simple form within the $|x,y,z\rangle$ basis set:
\begin{eqnarray}
\langle i |\hat S_{k} | j \rangle =i \epsilon _{ikj}.
\end{eqnarray}

We start by introducing the following set of S=1 coherent states characterized by vectors $\bf a$ and $\bf b$ satisfying the normalization constraint\,\cite{Nadya} 
\begin{eqnarray}
|{\bf c}\rangle =|{\bf a}, {\bf b}\rangle = {\bf c}\cdot{\bf \Psi}=({\bf a} +i{\bf b})\cdot{\bf \Psi} \, ,
\label{ab}
\end{eqnarray}
where ${\bf a}$ and ${\bf b}$ are real vectors that are
arbitrarily oriented with respect to some fixed coordinate
system in the pseudospin space with orthonormal basis ${\bf e}_{1,2,3}$. 

The two vectors are coupled, so that the minimal number of
dynamic variables describing the $S=1$ (pseudo)spin system appears to be equal to four.
Hereafter we would like to emphasize the $director$ nature of the ${\bf c}$
vector field: $|{\bf c}\rangle$ and $|-{\bf c}\rangle$ describe the physically
identical states.

It should be noted that in real space the $|{\bf c}\rangle$ state corresponds to a quantum on-site superposition 
\begin{eqnarray}
|{\bf c}\rangle = c_{-1}|0\rangle +c_0|1\rangle +c_{+1}|2\rangle \, .
\label{function}
\end{eqnarray}
Existence of such unconventional on-site superpositions is a principal point of the model.
Below instead of $\bf a$ and $\bf b$ we will make use of a pair of unit vectors $\bf m$ and $\bf n$, defined as follows\,\cite{Knig}: 
$$\bf a\,=\,\cos\varphi\,\,{\bf m} ;\,\, \bf b\,=\,\sin\varphi\,\,{\bf n}\,.
$$

 For the averages of the principal pseudospin operators we obtain
$$
\langle {\bf S} \rangle = \sin2\varphi[{\bf m} \times {\bf n}]\,;
$$
\begin{eqnarray}
\langle\{S_{i},S_{j}\}\rangle =2(\delta_{ij}-\cos^2\varphi \,m_{i}m_{j}-\sin^2\varphi \,n_{i}n_{j})\, ,
\end{eqnarray}
or
$$
\langle S_{i}^2\rangle =1-\frac{1}{2}(m_{i}^2+n_{i}^2)-\frac{1}{2}(m_{i}^2-n_{i}^2)\cos2\varphi \, , 
$$
$$
\langle\{S_{i},S_{j}\}\rangle =-(m_{i}m_{j}+n_{i}n_{j})-
$$
\begin{eqnarray}
(m_{i}m_{j}-n_{i}n_{j})\cos2\varphi\,, (i\not= j) \,.
\end{eqnarray}
\begin{figure*}[t]
\begin{center}
\includegraphics[width=14cm,angle=0]{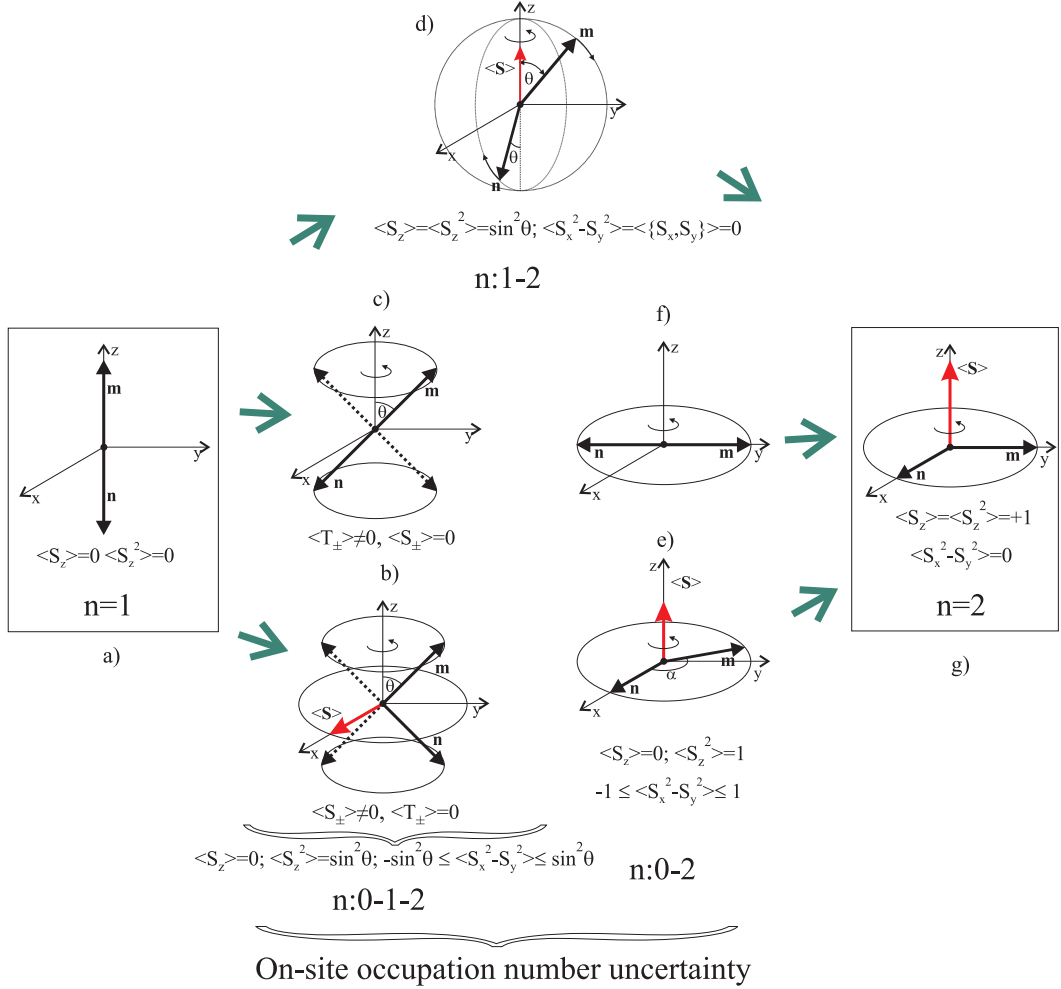}
\caption{(Color online) Cartoon showing orientations of the ${\bf m}$ and  ${\bf n}$ vectors which provide extremal values of different on-site pseudospin order parameters given $\varphi = \pi /4$ (see text for more detail).} \label{fig1}
\end{center}
\end{figure*}
One should note a principal difference between the $S=\frac{1}{2}$ and  $S=1$ quantum systems. The only on-site order parameter in the former case is an average spin moment $\langle S_{x,y,z}\rangle$, whereas in the latter one has five additional "spin-quadrupole", or spin-nematic order parameters described by traceless symmetric tensors
\begin{equation}
	Q_{ij}=\langle (\frac{1}{2}\{S_{i},S_{j}\}-\frac{2}{3}\delta_{ij})\rangle .
\end{equation}
 Interestingly, that in a sense, the  $S=\frac{1}{2}$ quantum spin system is closer to a classic one  ($S\rightarrow \infty$) with all the order parameters defined by a simple on-site vectorial order parameter $\langle {\bf S}\rangle$  than the  $S=1$ quantum spin system with its eight independent on-site  order parameters.

The operators $V^{k}_{q}$ ($q\not=0$) change the $z$-projection of the
pseudospin and transform the $|SM_S\rangle$ state into  the $|SM_{S}+q\rangle$
one. In other words, these can change the occupation number.
It should be emphasized that for the $S=1$ pseudospin algebra there are two operators: $V^{1}_{\pm 1}$ and $V^{2}_{\pm 1}$, or $S_{\pm}$ and $T_{\pm}=\{S_z, S_{\pm}\}$ that change the pseudo-spin projection (and occupation number) by $\pm 1$, with slightly different properties
\begin{equation}
\langle 0 |\hat S_{\pm} | \mp 1 \rangle = \langle \pm 1 |\hat S_{\pm} | 0
\rangle =\mp 1, \label{S1}
\end{equation}
but
\begin{equation}
\langle 0 |\hat T_{\pm}| \mp 1 \rangle = -\langle \pm 1 |(\hat T_{\pm}| 0 \rangle =+1. \label{S2}
\end{equation}
It is worth noting the similar behavior of the both operators under the hermitian conjugation:
${\hat S}_{\pm}^{\dagger}=-{\hat S}_{\mp}$;  ${\hat T}_{\pm}^{\dagger}=-{\hat T}_{\mp}$.

The $V^{2}_{\pm 2}$, or ${\hat S}_{\pm}^{2}$ 
operator changes the pseudo-spin projection by $\pm 2$ with the local order parameter 
$$
\langle S_{\pm}^{2} \rangle\,=\,\frac{1}{2}(\langle S_x^2-S_y^2\rangle \pm i\langle\{S_x,S_y\}\rangle )=
$$
\begin{equation}
c_+^*c_-=c_x^2-c_y^2\pm 2ic_xc_y\, .
\end{equation}
Obviously, this on-site off-diagonal order parameter is nonzero only when both $c_+$ and $c_-$ are nonzero, or for the on-site $0-2$ superpositions. It is worth noting that the ${\hat S}_{+}^{2}$ (${\hat S}_{-}^{2}$) operator
creates an on-site boson pair, or dimer, with a kinematic constraint $({\hat S}_{\pm}^{2})^2$\,=\,0, that underlines its "hard-core"\, nature.

Figure\,1 shows orientations of the ${\bf m}$ and  ${\bf n}$ vectors which provide extremal values of different on-site pseudospin order parameters given $\varphi = \pi /4$. The $n=1$ center is described by a pair of $\bf m$ and $\bf n$ vectors directed along Z-axis with $|m_z|=|n_z|$\,=\,1. We arrive at the  $1-2$ or $1-0$ mixtures if turn $c_{-1}$ or $c_{+1}$, respectively, into zero. The mixtures  are described by a pair of $\bf m$ and $\bf n$ vectors whose projections on the XY-plane, ${\bf m}_{\perp}$ and ${\bf n}_{\perp}$, are of the same length and orthogonal to each other: ${\bf m}_{\perp}\cdot{\bf n}_{\perp}$\,=\,0, $m_{\perp}$\,=\,$n_{\perp}$ with $[{\bf m}_{\perp}\times{\bf n}_{\perp}]$\,=\,$\langle S_z\rangle$\,=\,$\pm \sin^2\theta$ for $1-2$ and $1-0$ mixtures, respectively (see Fig.\,1). 

It is worth noting that for "conical"\, configurations in Figs.\,1b-1d:
$$
\langle S_z\rangle=0;\, \langle S_z^2\rangle=\sin^2\theta;\,\langle S_{\pm}^2\rangle=- \frac{1}{2}\sin^2\theta \,e^{\pm 2i\varphi}
$$
\begin{equation}
\langle S_{\pm}\rangle=-\frac{i}{\sqrt{2}}\sin2\theta \,e^{\pm i\varphi};\, \langle T_{\pm}\rangle=0 \, ,
\end{equation}
(Fig.\,1b);
$$
\langle S_z\rangle=0;\, \langle S_z^2\rangle=\sin^2\theta;\,\langle S_{\pm}^2\rangle=- \frac{1}{2}\sin^2\theta \,e^{\pm 2i\varphi}
$$
\begin{equation}
\langle S_{\pm}\rangle=0;\,\langle T_{\pm}\rangle=\mp \frac{1}{\sqrt{2}}\sin2\theta \,e^{\pm i\varphi};\, 
\end{equation}
(Fig.\,1c);
$$
\langle S_z\rangle=-\langle S_z^2\rangle=-\sin^2\theta;\,\langle S_{\pm}^2\rangle=0
$$
\begin{equation}
\langle S_{\pm}\rangle=\langle T_{\pm}\rangle=\pm \frac{1}{2}e^{\mp i\frac{\pi}{4}}\sin2\theta \,e^{\pm i\varphi}\, ,
\end{equation}
(Fig.\,1d). Figures 1e,f do show the orientation of ${\bf m}$ and ${\bf n}$ vectors for the local binary mixture $0-2$, and Fig.1g does for $n=2$ center.
It is worth noting that for binary mixtures $\left|1\right\rangle$-$\left|0\right\rangle$ and $\left|1\right\rangle$-$\left|2\right\rangle$  we arrive at the same algebra of the ${\hat S}_{\pm}$ and ${\hat T}_{\pm}$ operators with $\langle S_{\pm}\rangle=\langle T_{\pm}\rangle$, while for ternary mixtures $\left|0\right\rangle$-$\left|1\right\rangle$-$\left|2\right\rangle$ these operators describe different excitations.
Interestingly that in all the cases the local $n=1$ fraction can be written as follows:
\begin{equation}
	\rho (n=1)=1-\langle S_z^2\rangle = \cos^2\theta \, .
\end{equation}

In the bosonic language $\langle S_z\rangle$ and $\langle S_z^2\rangle$ are on-site diagonal order parameters, these describe 
local density and boson nematic order, respectively. The on-site mean values $\langle S_{\pm}\rangle$ and $\langle T_{\pm}\rangle$ 
are the two types of local off-diagonal order parameters that describe one-boson superfluidity, while $\langle S^2_{\pm}\rangle$ 
is a local order parameter of the two-boson, or dimer superfluidity.

\section{Effective S=1 pseudospin Hamiltonian}

General form of the effective pseudospin Hamiltonian which does commute with the z-component of the total pseudospin  $\sum_{i}S_{iz}$ thus conserving the mean boson density reads as follows\,\cite{Moskvin-11,JPCM}: 
$$
	{\hat H} =  \sum_{i}  (\Delta _{i}S_{iz}^2 - (\mu -h_{i})S_{iz})+ 
$$	
\begin{equation}	
	\sum_{k_1k_2q}\sum_{i<j} I_{k_1k_2q}(ij){\hat V}^{k_1}_{q}(i){\hat V}^{k_2}_{-q}(j)\, .
\end{equation}
The Hamiltonian can be rewritten to be a sum of potential and kinetic energies, that is of the $q=0$ ("diagonal") ${\hat H}_{ch}$ and $q\not= 0$ ("off-diagonal") ${\hat H}_{tr}$ terms: 
\begin{equation}
	{\hat H}={\hat H}_{ch}+{\hat H}_{tr}\,,
\label{H}	
\end{equation}
where
\begin{equation}
	{\hat H}_{ch} =  \sum_{i}  (\Delta _{i}S_{iz}^2
	  - (\mu -h_{i})S_{iz}) + \sum_{i<j} V_{ij}(S_{iz}S_{jz}+\alpha S_{iz}^2S_{jz}^2)\, ,
\label{Hch}	   
\end{equation}
and ${\hat H}_{tr}={\hat H}_{tr}^{(1)}+{\hat H}_{tr}^{(2)}$ being a sum of one-particle and two-particle transfer contributions
$$ 
{\hat H}_{tr}^{(1)}= \sum_{i<j} t^S_{ij}(S_{i+}S_{j-}+S_{i-}S_{j+})+
 \sum_{i<j} t_{ij}^{T}(T_{i+}T_{j-}+T_{i-}T_{j+})
$$ 
\begin{equation} 
 +\sum_{i<j} t_{ij}^{ST}(S_{i+}T_{j-}+S_{i-}T_{j+}+T_{i+}S_{j-}+T_{i-}S_{j+})\, ;
\label{H1}	
\end{equation}
\begin{equation}
  {\hat H}_{tr}^{(2)}=\sum_{i<j} t_{ij}^d(S_{i+}^{2}S_{j-}^{2}+S_{i-}^{2}S_{j+}^{2})\,,
  \label{H2}
\end{equation}
with a boson density constraint: 
\begin{equation}
\frac{1}{2N}\sum _{i} \langle S_{iz}\rangle =\Delta n	\, ,
\end{equation}
where $\Delta n$ is the deviation from a half-filling ($n=1$). 


Hamiltonian ${\hat H}_{ch}$ corresponds to a classical spin-1 Ising model with a single-ion anisotropy term, or the generalized Blume-Capel model\,\cite{Blume}, 
in the presence of a longitudinal magnetic field. 
The first single-site term in ${\hat H}_{ch}$ describes the effects of a bare pseudo-spin splitting and relates with the on-site density-density interactions: $\Delta$=$U$. The second term   may be
related to a   pseudo-magnetic field ${\bf h}_i$\,$\parallel$\,$Z$, which acts as a chemical potential ($\mu$ is the boson chemical potential, and $h_i$ is a (random) site energy). At variance with the real external field the chemical potential depends both on the parameters of the Hamiltonian (\ref{H}) and the temperature. 
 The third bilinear  and forth biquadratic terms in ${\hat H}_{ch}$ describe the effects of the short-  and long-range inter-site density-density interactions.

${\hat H}_{tr}$ plays the role of the kinetic energy where ${\hat H}_{tr}^{(1)}$ and ${\hat H}_{tr}^{(2)}$ describe the one-  and two-particle inter-site  hopping, respectively.
Hamiltonian ${\hat H}_{tr}^{(1)}$ represents an obvious extension of the conventional Hubbard model that assumes that the single particle orbital is infinitely rigid irrespective of occupation number, and has much in common with so-called dynamic Hubbard models\,\cite{Hirsch} that describe a correlated hopping. The ST and TT terms describe a density-dependent single-particle hopping.
It was Hirsch and coworkers\,\cite{Hirsch} who stressed the importance of the density-induced tunneling
effects in the condensed-matter context. 

However, before mapping the pseudospin model into a discrete
free bosonic model, one should take care that the amplitude of
the one-particle hoppings in Bose-Hubbard Hamiltonian (\ref{BH}) obey the bosonic commutation
relations. This implies that the amplitude of the $\left|1\right\rangle$$\left|2\right\rangle$-$\left|2\right\rangle$$\left|1\right\rangle$ process is twice as large as that of $\left|0\right\rangle$$\left|1\right\rangle$-$\left|1\right\rangle$$\left|0\right\rangle$ and a factor $\sqrt{2} $ larger than that of $\left|1\right\rangle$$\left|1\right\rangle$-$\left|2\right\rangle$$\left|0\right\rangle$.
It should be noted that within the triplet basis $\left|0,1,2\right\rangle$ the bosonic annihilation operator reads as follows\,\cite{Berg}
 $$
	{\hat b}_i=\frac{1}{2}\left[(1+\sqrt{2}){\hat S}_{i-}-(1-\sqrt{2}){\hat T}_{i-}\right]=
$$	
	\begin{equation}
	\sqrt{2}\left(1+\frac{\sqrt{2}-1}{\sqrt{2}}{\hat S}_{iz}\right){\hat S}_{i+} \, .
\end{equation}
In other words, in frames of a standard EBHM approach all the SS, TT, and ST terms in pseudospin kinetic energy are governed by the only Hubbard transfer integral $t$: 
$$
t^{S}=-\frac{(1+\sqrt{2})^2}{4}t;\, t^{T}=-\frac{(1-\sqrt{2})^2}{4}t;
$$
\begin{equation}
t^{ST}=-\sqrt{2}t;\,t^d=0 \, ,
\end{equation}
while the pseudospin Hamiltonian ${\hat H}_{tr}$ allows us to describe more complicated transfer mechanisms.
The one- and two-particle hopping terms in ${\hat H}_{tr}$ are of a primary importance for the transport properties of our model system, and deserve special attention.
Three (SS-, TT-, and ST-) types  of the one-particle hopping terms are
governed by the three transfer integrals $t^S_{ij}$, $t_{ij}^{T}$, $t_{ij}^{ST}$, respectively. Instead of ${\hat S}_{\pm}$ and ${\hat T}_{\pm}$ operators we can introduce two novel operators ${\hat P}_{\pm}$ and ${\hat N}_{\pm}$ as follows
\begin{equation}
	{\hat P}_{\pm}=\frac{1}{2}({\hat S}_{\pm}+{\hat T}_{\pm});\,{\hat N}_{\pm}=\frac{1}{2}({\hat S}_{\pm}-{\hat T}_{\pm})\, ,
\end{equation} 
so that the single particle transfer Hamiltonian transforms into 
$$ 
{\hat H}_{tr}^{(1)}= \sum_{i<j} t^P_{ij}(P_{i+}P_{j-}+P_{i-}P_{j+})+
 \sum_{i<j} t^N_{ij}(N_{i+}N_{j-}+N_{i-}N_{j+})
$$ 
\begin{equation} 
 +\sum_{i<j} t^{PN}_{ij}(P_{i+}N_{j-}+P_{i-}N_{j+}+N_{i+}P_{j-}+N_{i-}P_{j+})\, ;
\label{H1a}	
\end{equation}
where
\begin{equation}
	t^P_{ij}=t^S_{ij}+t_{ij}^{T}+t_{ij}^{ST};\, t^N_{ij}=t^S_{ij}+t_{ij}^{T}-t_{ij}^{ST};\,
	t^{PN}_{ij}=t^S_{ij}-t_{ij}^{T}\, .
\end{equation}
All the three terms here have a clear physical interpretation. The first $PP$-type term describes 
one-particle hopping processes $\left|1\right\rangle$$\left|2\right\rangle$-$\left|2\right\rangle$$\left|1\right\rangle$
that is  a rather conventional  motion of the extra boson in the lattice with the $n=1$ on-site occupation or the motion of the boson hole in the lattice with the $n=2$ on-site occupation. 
The second $NN$-type term describes one-particle hopping processes $\left|1\right\rangle$$\left|0\right\rangle$-$\left|0\right\rangle$$\left|1\right\rangle$ 
that is  a rather conventional  motion of a boson hole in the lattice with the $n=1$ on-site occupation  or the motion of a boson in the lattice with the $n=0$ on-site occupation.
These hopping processes are typical ones for heavily underfilled ($\left\langle n\right\rangle\ll 1$) or heavily overfilled ($\left\langle n\right\rangle \leq 2$) lattices, respectively. 
It is worth noting that the ST-type contribution of the one-particle transfer differs in sign for the $PP$ and $NN$ transfer thus breaking the ``particle-hole'' symmetry.

The third $PN$ ($NP$) term in (\ref{H1a}) defines a very different one-particle hopping process
$\left|1\right\rangle$$\left|1\right\rangle$-$\left|2\right\rangle$$\left|0\right\rangle$ ($\left|0\right\rangle$$\left|2\right\rangle$)
that is the particle-hole creation/annihilation. It should be noted that the ST-type transfer does not contribute to the reaction.

The two-particle(hole), or dimer hopping is governed by the transfer integral
  $t_{ij}^d$ that defines  a probability amplitude for the ``exchange'' reaction $\left|0\right\rangle$$\left|2\right\rangle$-$\left|2\right\rangle$$\left|0\right\rangle$, 
 or the   motion of an on-site dimer in the
lattice with the $n=0$ on-site occupation or  the   motion of an on-site hole $n=0$ in the
lattice with the $n=2$ on-site occupation.

All the kinetic energies can be rewritten in terms of the Cartesian pseudospin components, if we take into account that
$$
(S_{i+}S_{j-}+S_{i-}S_{j+})= -(S_{ix}S_{jx}+S_{iy}S_{jy})\,;
$$
$$
(S_{i+}S_{j-}-S_{i-}S_{j+})= i(S_{ix}S_{jy}-S_{iy}S_{jx})=i\left[{\bf S}_1\times{\bf S}_2\right]_z\,;
$$
$$ 
(T_{i+}T_{j-}+T_{i-}T_{j+})=-(T_{ix}T_{jx}+T_{iy}T_{jy})=
$$
$$
-(S_{ix}S_{jx}+S_{iy}S_{jy})S_{iz}S_{jz}-S_{iz}(S_{ix}S_{jx}+S_{iy}S_{jy})S_{jz}+h.c \,;
$$
$$ 
(T_{i+}T_{j-}-T_{i-}T_{j+})=i(T_{ix}T_{jy}-T_{iy}T_{jx})=i\left[{\bf T}_1\times{\bf T}_2\right]_z\, ;
$$.
$$
(S_{i+}T_{j-}+S_{i-}T_{j+})+h.c.=-\{(S_{iz}+S_{jz}),(S_{ix}S_{jx}+S_{iy}S_{jy})\}\,;
$$
$$
(S_{i+}^{2}S_{j-}^{2}+S_{i-}^{2}S_{j+}^{2})=
$$
$$
\frac{1}{2}\left[(S_{ix}^2-S_{iy}^2)(S_{jx}^2-S_{jy}^2)+\{S_{ix},S_{iy}\}\{S_{jx},S_{jy}\}\right]\, ;
$$
$$
(S_{i+}^{2}S_{j-}^{2}-S_{i-}^{2}S_{j+}^{2})=
$$
\begin{equation}
-\frac{i}{2}\left[(S_{ix}^2-S_{iy}^2)\{S_{jx},S_{jy}\}-\{S_{ix},S_{iy}\}(S_{jx}^2-S_{jy}^2)\right] \,.	
\end{equation}

Hamiltonian ${\hat H}_{ch}$ describes two types of a longitudinal long-range diagonal Z-ordering measured by the static structure factors such as
\begin{equation}
	S_{zz}({\bf q})=\frac{1}{N}\sum_{m,n}e^{-i{\bf q}\cdot ({\bf R}_m-{\bf R}_n)}\langle S_{mz}S_{nz}\rangle
\end{equation}
for a pseudospin-dipole order and 
\begin{equation}
	S^2_{zz}({\bf q})=\frac{1}{N}\sum_{m,n}e^{-i{\bf q}\cdot ({\bf R}_m-{\bf R}_n)}\langle S^2_{mz}S^2_{nz}\rangle\, ,
\end{equation}
for a pseudospin-quadrupole (nematic) order, respectively.

Hamiltonian ${\hat H}_{tr}$ describes different types of transverse long-range off-diagonal XY-ordering measured by the transverse components of the static structure factors such as
\begin{equation}
	S_{+-}({\bf q})=\frac{1}{N}\sum_{m,n}e^{-i{\bf q}\cdot ({\bf R}_m-{\bf R}_n)}\langle S_{m+}S_{n-}\rangle
\end{equation}
for conventional pseudospin-dipole order or
\begin{equation}
	T_{+-}({\bf q})=\frac{1}{N}\sum_{m,n}e^{-i{\bf q}\cdot ({\bf R}_m-{\bf R}_n)}\langle T_{m+}T_{n-}\rangle\, ,
\end{equation} and 
\begin{equation}
	S^2_{+-}({\bf q})=\frac{1}{N}\sum_{m,n}e^{-i{\bf q}\cdot ({\bf R}_m-{\bf R}_n)}\langle S^2_{m+}S^2_{n-}\rangle\, ,
\end{equation}
for two types of pseudospin-quadrupole (nematic) order. In conventional bosonic language the structure factors $S_{zz}({\bf q})$ and  $S^2_{zz}({\bf q})$ describe density-density correlations, the $S_{+-}({\bf q})$ and $T_{+-}({\bf q})$ do the single-boson superfluid correlations, while the $S^2_{+-}({\bf q})$ does the two-boson (on-site dimer) superfluid correlations.

\section{Typical simplified S=1 spin model}
Despite many simplifications, the
effective pseudospin Hamiltonian (\ref{H}) is rather complex, and represents one of the
most general forms of the anisotropic S=1 non-Heisenberg Hamiltonian.
Its real spin counterpart corresponds to an anisotropic S=1 magnet with a single-ion (on-site) and two-ion (inter-site bilinear and biquadratic) symmetric anisotropy in an external magnetic field under conservation of the total $S_z$.   
Spin Hamiltonian (\ref{H}) describes an interplay of the Zeeman,   single-ion and two-ion anisotropic terms giving rise to a competition of an (anti)ferromagnetic  order along Z-axis with an in-plane $XY$ magnetic order. Simplified versions of anisotropic S=1 Heisenberg Hamiltonian with bilinear exchange have been investigated rather extensively in recent years. Their analysis seems to provide an instructive introduction to description of our generalized pseudospin model.

 Typical S\,=\,1 spin Hamiltonian with uniaxial single-site and exchange anisotropies reads as follows:
$$
	{\hat H}=\sum_{i>j}J_{ij}(S_{ix}S_{jx}+S_{iy}S_{jy}+\lambda S_{iz}S_{jz})+
$$	
\begin{equation}	
	\sum_iDS_{iz}^2 - \sum_ihS_{iz}\, .
	\label{Hs1}
\end{equation}
Correspondence with our pseudospin Hamiltonian points to $D=\Delta$, $J_{ij}=t_{ij}$, $\lambda J_{ij}=V_{ij}$.
Usually one  considers the antiferromagnet with $J>0$ since, in general, this is the case of
more interest. However, the Hamiltonian (\ref{Hs1}) is invariant under the transformation $J,\lambda\rightarrow -J,-\lambda$ and a shift of the Brillouin zone ${\bf k}\rightarrow {\bf k}+(\pi ,\pi )$ for 2D square lattice. The system described by the Hamiltonian (\ref{Hs}) can be characterized by local (on-site) spin-linear order parameters $\langle {\bf S}\rangle$ and spin-quadratic (quadrupole spin-nematic) order parameters $Q^2_0=Q_{zz}=\langle S_z^2-\frac{2}{3}\rangle$ and $Q^2_{\pm 2}=\langle S_{\pm 1}^2\rangle$.

The model has been studied by several methods, e.g., molecular field approximation, spin-wave theories, exact numerical diagonalizations, nonlinear sigma model, quantum Monte Carlo, series expansions, variational methods, coupled cluster approach, self-consistent harmonic approximation,  and generalized SU(3) Schwinger boson representation\,\cite{Khajehpour,Sengupta1,Sengupta2,Oitmaa,Pires}.

The spectrum of the spin Hamiltonian (\ref{Hs1}) in the absence of external magnetic field changes drastically as $\Delta$
varies from very small to very large positive or negative values. A strong "easy-plane" anisotropy for large positive $\Delta >0$ favors a singlet phase where spins are in the $S_z = 0$   ground state. This ``quadrupole'' phase has no magnetic order, and is aptly referred to as a quantum paramagnetic phase (QPM), which is separated
from the "ordered" state by a quantum critical point at some $\Delta$\,=\,$\Delta_c^{QPM}$. This is a quadrupole state with no magnetic order, so that all linear order parameters vanish and only a quadrupole (spin-nematic) order parameter such as $Q_{zz}=\langle S_z^2-\frac{2}{3}\rangle$ is nonzero.
The QPM  phase consists of a unique ground state with total spin $S_z^{total}$\,=\,0, separated by a gap from the first excited states, which lie in the sectors $S_z^{total}=\pm 1$. 
It is worth noting that the QPM order differs in principle from the conventional paramagnetic state, because for S\,=\,1 in the classical paramagnetic state $\langle S_x^2\rangle$\,=\,$\langle S_y^2\rangle$\,=\,$\langle S_z^2\rangle$\,=\,2/3, while in the quantum paramagnetic state $\langle S_z^2\rangle$\,=\,0, $\langle S_x^2\rangle$\,=\,$\langle S_y^2\rangle$\,=\,1.
Strictly speaking, all the above analysis concerns the typical mean-field approximation (MFA). Beyond the MFA the QPM ground state 
contains an admixture of states formed by exciton-like tightly bound particle-hole fluctuations ($0-2$ on nearby sites).

A strong "easy-axis" anisotropy for large negative $\Delta \leq\Delta_c^{IS}$, $\Delta_c^{IS}$\,=\,$2(V_{nn}/t_{nn}-1)$\,\cite{Pires}, favors a spin ordering along $z$, the "easy axis", with the on-site $S_z = \pm 1$ ($Z$-phase). The order parameter will be "Ising-like"\, and long-range (staggered) diagonal order will persist at finite temperature, up to a critical line T$_c(\Delta )$. The easy axis antiferromagnetic  $Z_{AFM}$ phase or more complicated long-range spin $Z$-order  are characterized by  the longitudinal component of the static structure factor $S_{zz}({\bf q})$.


For intermediate values $\Delta_c^{QPM}>\Delta >\Delta_c^{IS}$ the system is in a gapless XY phase where the spins will be preferentially in the $xy$ plane (choosing $z$ as the hard axis) and the Hamiltonian will have O(2) symmetry. At T = 0 this symmetry will be spontaneously broken and the system will exhibit spin order in some direction, reduced by quantum fluctuations. The broken O(2) symmetry will result in a single gapless Goldstone mode.  Although there will be no ordered phase at finite temperature one expects a finite temperature Kosterlitz-Thouless transition.
The XY phase has long-range off-diagonal ordering measured by the transverse component of the static structure factor $S_{+-}({\bf q})$.

 For large positive $\Delta$, in the QPM phase, the low energy excitations arise from exciting one of the $S_z = 0$ ($n=1$) sites to $S_z = +1$ ($n=2$)  or $S_z = -1$ ($n=0$). Such a local excitation, actually the effective particle or hole, can then propagate through the lattice  due to the transfer terms (quantum fluctuations) in the $H_{tr}$, forming  a well defined quasiparticle (magnon) band  with energy $\varepsilon ({\bf k})$. These coherent magnon bands will have an energy gap, which we expect will vanish as $\Delta$\,$\rightarrow$\,$\Delta_c^{QPM}$. An analytic expression for  $\varepsilon ({\bf k})$ in the QPM phase has been proposed by Papanicolaou\,\cite{Papanicolaou}, based on a generalized Holstein-Primakoff transformation for isotropic $nn$-Heisenberg model with the single-site anisotropy. 
The application of an effective field, $h_z$, along the z-axis reduces the spin gap
linearly in $h_z$ since the field couples to a conserved quantity
(total spin along the z-axis). The gap is closed at a critical field $h_c$ (the quantum critical point (QCP)) where the bottom of the $S_z$\,=\,1 branch of (pseudo)spin excitations touches zero. This QCP belongs to the BEC universality class and the gapless mode of low-energy $S_z$\,=\,1 excitations remains quadratic for small momenta, because the Zeeman term commutes with the rest of the Hamiltonian.

 Both excitation branches in the QPM phase, $\Delta S_z$\,=\,$\pm 1$ (particle/hole) have the same dispersion at zero field, $h_z=0$,
as expected from time reversal symmetry. A finite $h_z$ splits
the branches linearly in $h_z$: $\varepsilon_{\pm} ({\bf k})\rightarrow\varepsilon_{\pm} ({\bf k})\pm h_z$ without changing the dispersion.
This is a consequence of the fact that the external field couples to the total spin $\sum S_z$, which is a conserved quantity. 

It should be noted that there are three types of two-magnon excitations with $\Delta S_z^{total}$\,=\,+2,\,-2, and 0, respectively. 
Two-magnon bound state with $\Delta S_z^{total}$\,=\,0, or coupled particle-hole pair can propagate through the lattice, forming a quasiparticle band.  
   
At least for relatively small negative $\Delta <\Delta_c^{IS}$  the lowest energy excitations, in the unperturbed system, consist of a single spin excited from its ordered $S_z = \pm 1$ state to $S_z$\,=\,0, i.e. $\Delta S_z =\mp 1$.  Respective coherent magnon band will have an energy gap at $\Gamma$ point $(0,0)$, which behaves like $\varepsilon (0,0)\sim 2\sqrt{2V_{nn}|\Delta |}$ at small $|\Delta |$. This reflects, in the easy axis case, the fact that the remnant O(2) symmetry of the Hamiltonian is not spontaneously broken in this case, and so Goldstone modes are absent.

However, for large negative $\Delta$ the single-magnon (single-particle) excitations will not be the lowest energy excitations of the system. Their energy will be of
order $\left|\Delta\right|$, whereas an excitation with $\Delta S_z =\pm 2$ (i.e. $S_z = \pm 1 \leftrightarrow S_z = \mp 1$) will have an energy of order $2zV_{nn}$ as $\Delta\rightarrow -\infty $. Such a two-particle (local dimer) excitation, created at a particular site, can again propagate through the lattice, forming a quasiparticle band. We may think of this local dimer as a long-lived virtual two-magnon bound state (bimagnon) where the magnons are bound on the same site.

 
 Hamer {\it et al.}\,\cite{Oitmaa} have shown that at finite effective field $h_z$ but $\lambda$\,=\,1 the XY phase transforms into a canted antiferromagnetic XY-Z$_{FM}$ phase which appears right above $h_c$: the spins acquire a uniform longitudinal component and an antiferromagnetically ordered transverse component that spontaneously breaks the U(1) symmetry of global spin rotations along the z-axis. The longitudinal magnetization  increases with field and saturates at the fully polarized
(FP) state (all $S_z$\,=\,1) above the saturation field $h_s$. The FP state corresponds to a bosonic Mott insulator in the language of Bose gases.  
 
 The  field induced quantum phase transition from the QPM to the XY-Z$_{FM}$ phase is
qualitatively different from the transition between the same two phases that is induced by a change of $\Delta$ at $h_z$\,=\,0. If the single-ion anisotropy is continuously
decreased at zero applied field, the two excitation branches remain degenerate and the gap vanishes at $\Delta = \Delta_c^{QPM}$ ($h_z$\,=\,0). The low energy dispersion becomes linear at the QPM-CAFM phase boundary  for small k. However, the degeneracy between the two branches at $h_z$\,=\,0 is lifted inside the CAFM phase – one of the branches remains gapless with a linear dispersion at low energy (corresponding to the Goldstone mode of the ordered CAFM state )
whereas the other mode develops a gap to the lowest excitation.
the effect of increasing $h_z$ from zero at
a fixed $\Delta > \Delta_c^{QPM}$ is to reduce the gap linearly in $h_z$ with no change of dispersion.

At $D>$\,0  and $\lambda >$\,1 the phase diagram of the S\,=\,1 Heisenberg model  with
uniaxial anisotropy (\ref{Hs1}) contains an extended spin supersolid (SS) or “biconical” phase XY-Z$_{FIM}$ with a ferrimagnetic $z$-order that does exist over a range of magnetic fields. The model also exhibits other interesting phenomena such as magnetization plateaus and a multicritical point\,\cite{Sengupta1}. The magnetization stays zero up to the critical field, $h_{c1}$, that marks a quantum phase transition (QPT) to a state with a finite fraction of spins in all the $S_z$\,=\,0,$\pm$1 states. This spin supersolid state has a finite $S_{zz}(\pi ,\pi )$ as well as finite $S_{+-}(0,0)$.
The magnetization increases continuously up to $m_z$\,=\,0.5 at $h_{c2}$, where there is a second QPT to a second
Ising-like state (IS2) where all the $S_z$\,=\,-1 ($n=0$) sites have been flipped to the $S_z$\,=\,0 ($n=1$) state. At this, $S_{zz}(\pi ,\pi )$ remains divergent but $S_{+-}(0,0)$ drops to zero. Upon further increasing the field, there is a first order transition to a pure XY-AFM phase (CAFM) with the vanishing diagonal order but finite $S_{+-}(0,0)$. 
This situation persists until all the spins have flipped to the $S_z$\,=\,+1 ($n=2$) state (fully polarized, FP phase). The extent of the SS phase decreases with decreasing  $\lambda $ and vanishes for $\lambda\approx$\,1 leaving a second order transition from the SS to the XY (CAFM) phase.

At $D<$\,0, $J>$\,0,  and $\lambda =$\,1 the ground state of the spin Hamiltonian (\ref{Hs1}) corresponds to the easy axis antiferromagnetic  $Z_{AFM}$ phase. At small anisotropy, $\left|D\right|/J\leq 1$, the application of an effective field, $h_z$, along the z-axis induces first a rather conventional spin-flop transition to a pure XY-AFM phase (CAFM) with the vanishing diagonal order but finite $S_{+-}(0,0)$ ending by the transition to fully polarized ferromagnetic $Z_{FM}$ phase. However, at large anisotropy $\left|D\right|/J\gg 1$ instead of the mean-field first-type (metamagnetic) phase transition $Z_{AFM}$-$Z_{FM}$ we arrive at an unconventional intermediate phase with spin ferronematic (FNM) order characterized by zero value of the $S_{+-}(0,0)$ factor but nonzero $S^2_{+-}(0,0)$ correlation function\,\cite{Sengupta2}.

The phase diagram in the most interesting intermediate regime can change drastically, if we take into account  frustrative effects of next-nearest neighbor couplings or different non-Heisenberg biquadratic interactions\,\cite{Pires}.    
It should be noted that even for simple isotropic 2D-$nnn$ antiferromagnetic Heisenberg model the classical ground state has a N\'{e}el order only when $J_2/J_1<1/2$, where $J_1$ is the nearest-neighbor and $J_2$ the next-nearest neighbor interaction. However when $J_2/J_1>1/2$, the ground state consists of two independent sublattices with antiferromagnetic order. The
classical ground state energy does not depend on the relative orientations of both sublattices. However, quantum fluctuations lift this degeneracy and select a collinear order state, where the neighboring spins align ferromagnetically along one axis of the
square lattice and antiferromagnetically along the other (stripe-like order).

Turning to spin-boson mapping we note that QPM phase ($n_i=1$), fully polarized $Z_{FM}$ phases with $n_i=0$ or $n_i=2$ correspond to Mott insulating phases, the XY and XY-Z$_{FM}$ orderings correspond to a Bose-Einstein condensate (BEC)
of single bosons while FNM phase corresponds a  BEC of the boson dimers. The XY-Z$_{FIM}$ phases correspond to supersolids.

The pseudospin Hamiltonian, Eqs.(\ref{H})-(\ref{H2})  differs  from its  simplified version (\ref{Hs}) in several points.
First, this concerns the density constraint. It is worth noting that the charge density constraint in an uniform pseudospin system can be fulfilled only under some  quasidegeneracy.  
Second, the pseudospin parameters, in particular $\Delta$, $V_{ij}$, $h$ in the effective Hamiltonian (\ref{H}) can be closely linked to each other. Instead of a simple usually antiferromagnetic XY-exchange term in (\ref{Hs}) we should proceed with a significantly more complicated form of the "transversal"\, term in the pseudospin Hamiltonian, (\ref{H}), with inclusion of two biquadratic terms and unconventional "mixed" asymmetric  ST-type term which formally breaks the time inversion symmetry and is absent for conventional spin Hamiltonians. The seemingly main bilinear XY-exchange term in ${\hat H}_{tr}$ appears to be of the ferromagnetic sign. Along with a simple spin-linear planar XY-mode  with nonzero $\langle
S_{\pm}\rangle$ we arrive at two  novel spin-quadrupole nematic modes with nonzero $\langle T_{\pm}\rangle$ and/or $\langle S_{\pm}^2\rangle$. Hereafter we will denote different counterparts of the phases of the simple model (\ref{Hs}) as follows: novel XY-phase, Z$_{AFM}$ for Ising-type antiferromagnetic order along $z$-axis, XY-Z$_{FIM}$ for spin supersolid phases with simultaneous XY- and ferrimagnetic  orderings along $z$-axis, XY-Z$_{FM}$ for a phase with simultaneous XY- and ferromagnetic orderings along $z$-axis (the analogue of CAFM phase), and Z$_{FM}$ for fully $z$-polarized ferromagnetic phase.

\section{Topological defects in 2D S=1 pseudospin  systems}
\subsection{Short overview}
In the framework of our model the 2D Bose-Hubbard system prove to be in the universality    class of the (pseudo)spin 2D systems whose description incorporates static or dynamic     topological defects to be natural element both of micro- and macroscopic physics. 
Depending on the structure of effective pseudo-spin Hamiltonian in 2D-systems the latter could correspond to either in-plane and out-of-plane vortices or skyrmions. Under certain conditions either topological defects could determine the structure of the ground state. In particular, this could be a generic feature of electric multipolar systems with long-range multipolar interactions. Indeed, a Monte-Carlo simulation of a ferromagnetic Heisenberg model with dipolar interaction on a 2D square lattice $L\times L$
 shows that, as $L$ is increased, the spin structure changes from a ferromagnetic one to a novel one with a vortex-like arrangement of spins even for rather small magnitude of dipolar anisotropy\,\cite{Sasaki}.






Topological defects are stable non-uniform spin structures with broken translational symmetry and non-zero topological charge (chirality, vorticity, winding number). Vortices are stable states of anisotropic 2D Heisenberg Hamiltonian
\begin{equation}
	{\hat H}=\sum_{i>j}J_{ij}(S_{ix}S_{jx}+S_{iy}S_{jy}+\lambda S_{iz}S_{jz}) \, ,
	\label{H11}
\end{equation}
with the "easy-plane" anisotropy when the anisotropy parameter  $\lambda <1$.
Classical in-plane vortex ($S_{z}=0$) appears to be a stable solution of classical
Hamiltonian (\ref{H11}) at $\lambda  < \lambda _{c}$ ($\lambda _{c} \approx 0.7$
for square lattice). At  $1 > \lambda > \lambda _{c}$ stable solution corresponds
to the out-of-plane OP-vortex ($S_{z}\not= 0$), at which center the spin vector appears to be oriented along $z$-axis, and at infinity it arranges within $xy$-plane.  The in-plane vortex
is described by the formulas $\Phi=q\varphi$, $\cos\theta=0$. The $\theta(r)$
dependence  for the out-of-plane vortex cannot be found analytically. Both kinds of vortices have the energy logarithmically dependent on the size of the system.


The cylindrical domains, or bubble  like solitons with spins oriented  along the $z$-axis both at infinity and in the center (naturally, in opposite directions), exist for the "easy-axis" anisotropy $\lambda >1$. Their energy has a finite value.
 Skyrmions are general static solutions of classical continuous limit of the isotropic ($\lambda =1$) 2D Heisenberg ferromagnet, obtained by Belavin and Polyakov \cite{BP} from classical nonlinear sigma model. 
 Belavin-Polyakov skyrmion and out-of-plane vortex represent the simplest toy
model (pseudo)spin textures\,\cite{BP,Borisov}. 
 

The simplest skyrmion spin texture
looks like  a bubble domain in ferromagnet and consists of a vortex-like
arrangement of the in-plane components of spin with the $z$-component reversed
in the centre of the skyrmion and gradually increasing to match the homogeneous
background at infinity. 
 The spin distribution within such a classical skyrmion with a topological charge $q$ is given as follows\,\cite{BP}
\begin{equation}
\Phi=q\varphi +\varphi
_0;\quad\cos\Theta=\frac{r^{2q}-\lambda^{2q}}{r^{2q}+\lambda^{2q}}, \label{sk}
\end{equation}
where $r,  \varphi$ are polar coordinates on plane,  $q=\pm 1,\,\pm 2,...$ the chirality. For $q=1$, $\varphi_0$\,=\,0 we arrive at
\begin{equation}
n_{x}=\frac{2r\lambda }{r^{2}+\lambda ^{2}}\cos\varphi ;\,
n_{y}=\frac{2r\lambda }{r^{2}+\lambda ^{2}}\sin\varphi ;\,
n_{z}=\frac{r^{2}-\lambda ^{2}}{r^{2}+\lambda ^{2}}\, ,
\label{sk1}
\end{equation}
In terms of the stereographic variables the skyrmion with radius  $\lambda $   and phase  $\varphi _{0}$  centered at a point $z_0$ is identified with spin distribution  $w(z)=\frac{\Lambda }{z-z_0}$, where $z=x+iy=re^{i \varphi }$  is a point in the complex plane, $\Lambda =\lambda e^{i\alpha}$.
For a multicenter skyrmion we have\,\cite{BP}
\begin{equation}
w(z)=\cot\frac{\Theta}{2}\,e^{i\Phi}=\prod_i\left(\frac{z-z_j}{\Lambda}\right)^{m_j}\prod_j\left(\frac{\Lambda}{z-z_j}\right)^{n_j}\, ,
\end{equation}
where
 $\sum m_i >\sum n_j$, $q=\sum m_j$.
Skyrmions are characterized by the magnitude and sign of its topological
charge, by its size (radius), and by the global orientation of the spin. The
scale invariance of skyrmionic solution reflects in that its energy $E_{sk}=4\pi |q|IS^2$ is proportional to  topological charge and does not depend on  radius and global phase\,\cite{BP}. Like domain
walls,  vortices and skyrmions are stable for topological reasons. Skyrmions cannot decay into other configurations because of this topological stability no matter
how close they are in energy to any other configuration.

In a continuous field model, such as, e.g., the nonlinear 
$\sigma$-model, the ground-state energy of the skyrmion does not
depend on its size\,\cite{BP}, however, for  the skyrmion  on
a lattice, the energy depends on its size. This must
lead to the collapse of the skyrmion, making
it unstable.
Strong anisotropic interactions, in particular, long 
range dipole-dipole interactions may,
in principle, dynamically stabilize the skyrmions in 2D
lattices\,\cite{Abanov}.


Wave function of the spin system, which corresponds to a classical skyrmion,  is a product of spin coherent states \cite{Perelomov}. In case of spin  $S=\frac{1}{2}$
\begin{equation}
\Psi _{sk}( 0) =\prod\limits_{i}\lbrack \cos \frac{\theta _{i}}{2}e^{i\frac{\varphi _{i}}{2}}\mid \uparrow \rangle +\sin \frac{\theta _{i}}{2}e^{-i\frac{\varphi _{i}}{2}}\mid \downarrow \rangle \rbrack ,
\end{equation}
where $\theta _{i}=\arccos \frac{r_{i}^{2}-\lambda ^{2}}{r_{i}^{2}+\lambda ^{2}}$. Coherent state provides a maximal equivalence to classical state with minimal uncertainty of spin components.
The motion of such skyrmions has to be of highly quantum mechanical nature. However, this may involve a semi-classical percolation in the case of heavy non-localized skyrmions or variable range hopping in the case of highly localized skyrmions in a random potential.
Effective overlap and transfer integrals for quantum skyrmions are calculated analytically by Istomin and Moskvin \cite{Istomin}. The skyrmion motion has a cyclotronic character and resembles that of electron in a magnetic field.

The interest in skyrmions in ordered spin systems received much attention soon after the discovery of high-temperature superconductivity in copper oxides\,\cite{skyrmion-cuprates,skyrmion}. Initially, there was some hope that interaction of electrons and holes with spin skyrmions could play some role in superconductivity, but this was never successfully demonstrated. Some indirect evidence of skyrmions in the magnetoresistance of the litium doped lanthanum copper oxide has been recently reported\,\cite{skyrmion-LaLiCuO} but direct observation of skyrmions in 2D antiferromagnetic  lattices is still lacking.
In recent years the skyrmions and exotic skyrmion crystal (SkX) phases have been discussed in connection with a
wide range of condensed matter systems including quantum Hall effect, spinor Bose condensates and especially chiral magnets\,\cite{Bogdanov}.
It is worth noting that the skyrmion-like structures for hard-core 2D boson system were considered by Moskvin {\it et al.}\,\cite{bubble} in frames of the s=1/2 pseudospin formalism.


\subsection{Unconventional skyrmions in $S=1$ (pseudo)spin systems}

Different skyrmion-like topological defects for 2D (pseudo)spin S=1 systems as solutions of isotropic spin Hamiltonians were addressed in Ref\,\onlinecite{Knig}   and in more detail in Ref.\,\onlinecite{Nadya}. 
In general, isotropic non-Heisenberg spin-Hamiltonian for the $S=1$ quantum
 (pseudo)spin systems should include both bilinear Heisenberg exchange term and
 biquadratic non-Heisenberg exchange term:
\begin{equation}
	\hat{H}=-\tilde J_1\sum_{i,\eta}\hat{\bf{S}}_i
	\hat{\bf{S}}_{i+\eta}-\tilde
	J_2\sum_{i,\eta}(\hat{\bf{S}}_i\hat{\bf{S}}_{i+\eta})^2=
	\label{ha1} 
\end{equation}
$$=-J_1\sum_{i,\eta}\hat{\bf{S}}_i
\hat{\bf{S}}_{i+\eta}-J_2\sum_{i,\eta}\sum_{k\geq
j}^3(\{\hat{S}_k\hat{S}_j\}_i\{\hat{S}_k\hat{S}_j\}_{i+\eta})
$$
where $J_i$ are the appropriate exchange integrals, $J_1=\tilde
J_1-\tilde J_2/2$, $J_2=\tilde J_2/2$, $i$ and $\eta$ denote the
summation over lattice sites and nearest neighbours, respectively.


Having  substituted  our trial wave function (\ref{ab}) to $\langle {\hat
H}\rangle$ provided $\langle \hat{\bf S}(1)\hat{\bf S}(2)\rangle =\langle
\hat{\bf S}(1)\rangle \langle \hat{\bf S}(2)\rangle $ we arrive at
 the Hamiltonian of the isotropic classical spin-1 model in the continual
  approximation as follows:
$$
	H=J_1\int d^2\bf
	r\left[\sum_{i=1}^{3}(\vec{\nabla}\langle S_i\rangle )^2\right]+
$$	
\begin{equation}	
	J_2\int d^2{\bf
	r}\left[\sum_{i\geq
	j=1}^{3}(\vec{\nabla}a_ia_j+\vec{\nabla}b_ib_j)^2\right]+\frac{4(J_2-J_1)}{c^2}\int |\langle \hat{\bf S}\rangle|^2d^2{\bf r} \, , 
	\label{ham7}
\end{equation}
where $\langle \hat{\bf S}\rangle=2[\bf a\times\bf b]$. It should be noted that  the third
"gradientless" term in the Hamiltonian breaks the scaling invariance of the model. 

\subsubsection{Dipole (pseudo)spin skyrmions}

Dipole, or magnetic skyrmions as the solutions of bilinear Heisenberg
 (pseudo)spin Hamiltonian when $J_2=0$ were obtained in Ref.\,\onlinecite{Knig} given the restriction
   $\bf a\perp\bf b$ and the lengths of these vectors were fixed.

The model  reduces to the nonlinear
 O(3)-model with the solutions for $\bf a$ and $\bf b$  described by the following formulas
 (in polar coordinates):
$$
\sqrt{2}{\bf a}=({\bf e_z}\sin\theta-{\bf e_{r}}\cos\theta)\sin\varphi+
{\bf e_{\varphi}}\cos\varphi \, ;
$$
\begin{equation}
	\sqrt{2}{\bf b}=({\bf e_z}\sin\theta-{\bf e_{r}}\cos\theta)\cos\varphi-
	{\bf e_{\varphi}}\sin\varphi \,.
	\label{knig}
\end{equation}

For dipole "magneto-electric" skyrmions the $\bf m,\bf n$ vectors are assumed to be perpendicular to each other ($\bf m\perp \bf n$) and the (pseudo)spin structure is determined by the skyrmionic distribution (\ref{sk}) of the ${\bf l}=[{\bf m}\times{\bf n}]$ vector\,\cite{Knig}. In other words, the fixed-length spin vector $\langle {\bf S}\rangle =2[{\bf a}\times{\bf b}]$  is distributed in the same way as in the usual skyrmions (\ref{sk}).
However,
unlike the usual classic skyrmions, the dipole skyrmions in the S=1 theory 
have additional topological structure due to
 the existence of two vectors $\bf m$ and $\bf n$. Going around the center
 of the skyrmion the vectors can make $N$ turns around the
 ${\bf l}$ vector. Thus, we can introduce two topological
 quantum numbers: $N$ and $q$\,\cite{Knig}. In addition, it should be noted that
 $q$  number  may be half-integer.
 The dipole-quadrupole skyrmion is characterized by nonzero both pseudospin dipole order parameter $\langle \bf S\rangle$ with usual skyrmion texture (\ref{sk}) and quadrupole order parameters 
\begin{equation}
	\langle \{{\hat S}_{i}{\hat S}_{j}\}\rangle =2\langle {\hat S}_{i}\rangle \langle {\hat S}_{j}\rangle = l_il_j \, .
\end{equation}

\subsubsection{Quadrupole (pseudo)spin skyrmions}

Hereafter we address another situation with purely biquadratic
  (pseudo)spin Hamiltonian ($J_1$=0) and treat the
    non-magnetic (``electric'') degrees of freedom.
    The topological classification of the purely electric
    solutions is simple because it is also based on the usage of subgroup
     instead of the full group. We address the solutions given
      $\vec a\parallel\vec b$ and the fixed
      lengths of the vectors, so we use for the
     classification the same subgroup as above.

  After simple algebra the biquadratic part of the Hamiltonian can be reduced to the expression  familiar for nonlinear O(3)-model:
$$
	H_{bq}=J_2\int d^2{\bf r}\left[ \sum_{i,j=1}^{3}(\vec{\nabla}n_in_j)^2\right] =
$$	 
	\begin{equation}
	2J_2|{\bf n}|^{2}\int d^2{\bf r}\left[\sum_{i=1}^{3}(\vec{\nabla}n_i)^2\right]\, .
	\label{hm4} 
\end{equation}
where
${\bf a}=\alpha {\bf n}, {\bf b}=\beta {\bf n}$, and $\alpha+i\beta= \exp (i\kappa)$,
$\kappa\in R$, $|{\bf n}|^2=$const. 
Its solutions are skyrmions, but instead of
the spin distribution in magnetic skyrmion we have  solutions with zero spin,
 but the non-zero distribution of five spin-quadrupole moments $Q_{ij}$, or
   $\langle \{S_{i}S_{j}\}\rangle$ which in turn are determined
by the "skyrmionic"\, distribution of the ${\bf n}$ vector (\ref{sk})
with classical skyrmion energy: $E_{el}=16\pi qJ_2$. The distribution of the spin-quadrupole
moments $\langle \{S_{i}S_{j}\}\rangle$ can be easily obtained:
$$
\langle S_z^2\rangle =\frac{4r^{2q}\lambda^{2q}}{(r^{2q}+\lambda^{2q})^2}\,;
 \langle {\hat S}_{\pm}^2\rangle
=\frac{2r^{2q}\lambda^{2q}}{(r^{2q}+\lambda^{2q})^2}e^{\pm 2iq\varphi}\,;
$$
\begin{equation}
\langle {\hat T}_{\pm}\rangle =-i\sqrt{2}\frac{(\lambda^{2q}-r^{2q})r^q\lambda^q}{(r^{2q}+\lambda^{2q})^2}e^{\mp i
q\varphi}\,
\label{qq}
\end{equation}
  One should be emphasized
that the distribution of five independent quadrupole order parameters for the
quadrupole  skyrmion are straightforwardly determined by a single vector field
${\bf m}({\bf r})$ (${\bf n}({\bf r})$) while $\langle \hat {\bf S}\rangle$\,=\,0.
\begin{figure}[t]
\includegraphics[width=8.5cm,angle=0]{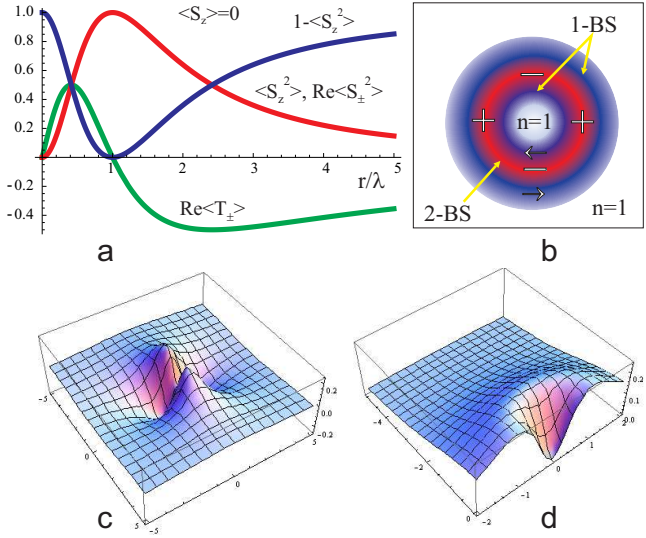}
\caption{(Color online) a) Radial distribution of the boson nematic order parameters for a quadrupole pseudospin skyrmion (q=1) with $\left\langle n_i\right\rangle =n=1$ ($\varphi =0$): b) the ring shaped distribution of the one- and two-boson SF order parameters: c) and d) the spatial distribution of $Re\langle {\hat S}_{\pm}^2\rangle $ and  $\langle {\hat S}_{z}^2\rangle $, respectively.}
\label{fig2}
\end{figure}
Fig.\ref{fig2} demonstrates the radial distribution of different (pseudo)spin order parameters for the quadrupole skyrmion. We see a circular layered structure with clearly visible anticorrelation effects due to a (pseudo)spin kinematics. Interestingly, at the center ($r=0$) and far from the center ($r\rightarrow\infty$) for such a skyrmion we deal with a $M$\,=\,0, or Mott insulating state while for the domain wall center ($r=\lambda$) we arrive at a $M$\,=\,$\pm$1 superposition with maximal value of the $|\langle {\hat S}_{\pm}^2\rangle |$ parameter whose weight diminishes with  moving away from the center. The $|\langle {\hat T}_{\pm}\rangle |$ parameter turns into zero at the domain wall center $r=\lambda$, at the skyrmion center $r=0$ and at  the infinity $r\rightarrow \infty$ ($\propto\frac{1}{r}$), with the two extremes at  $r=\frac{\lambda}{\sqrt{2} \pm 1}$. In other words, we arrive at a very complicated interplay of single and two boson superfluids with density maxima at $r=\frac{\lambda}{\sqrt{2} \pm 1}$ and at the domain wall center ($r=\lambda$), respectively. 
The ring shaped domain wall is an area with a circular distribution of the superfluid order parameters, or circular "bosonic" supercurrent. Nonzero $T$-type order parameter distribution points to a circular "one-boson" current with a puzzlingly opposite sign ($\pi$ phase difference) of the $\langle {\hat T}_{\pm}\rangle $ parameter for "internal"\, ($0<r<\lambda$) and "external"\, ($r>\lambda$) parts of the skyrmion, while the $\langle {\hat S}_{\pm}^2\rangle $ parameter defines the two-boson, or dimer superfluid order.
The specific spatial separation of different order parameters that avoid each other reflects the competition of different $k,j$ terms in (\ref{ha1}). 
Given the simplest winding number $q=1$ we arrive at the $p$ or $d$-wave 
($d_{x^2 - y^2}$/$d_{xy}$ in-plane symmetry of the one-boson or dimer superfluid order parameters, respectively. 

One of the most exciting features of the quadrupole skyrmion is that such a skyrmionic structure is characterized by an uniform distribution of the mean on-site boson density $\langle n_{i}\rangle$\,=\,$n$\,=1 as $\langle {\hat S}_{iz}\rangle $\,=\,0. In other words, the quadrupole skyrmionic structure and bare ``parent'' Mott insulating phase have absolutely the same distribution of the mean on-site densities. From the one hand, this point underlines an unconventional quantum nature of the quadrupole skyrmion under consideration, while from the other hand it makes the quadrupole skyrmion texture to be an ``invisible being'' for several experimental techniques. However, the domain wall center of the quadrupole skyrmion appears to reveal maximal values of the pseudospin susceptibility $\chi_{zz}$\,\cite{bubble}. It means the domain wall  appears to form a very efficient ring-shaped potential well for the boson localization thus giving rise to a novel type of a ``charged'' topological defect. In the framework of the pseudospin formalism the ``charging'' of a bare ``neutral'' skyrmion corresponds to a single-magnon $\Delta S_z$\,=\,$\pm$\,1 (single particle) or a two-magnon $\Delta S_z$\,=\,$\pm$\,2 (two-particle) dimer excitations. It is worth noting that for large negative $\Delta$ the single-magnon (single-particle) excitations may not be the lowest energy excitations of the strongly anisotropic pseudospin system. Their energy may  surpass the energy of a two-magnon bound state (bimagnon), or two-boson dimer excitation, created at a particular site. Thus we arrive at a competition of two types of ``charged'' quadrupole skyrmions with $\Delta N$\,=\,$\pm$\,1 and $\Delta N$\,=\,$\pm$\,2, respectively ($\Delta N$ is a total number of bosons). Such a ``charged'' topological defect can be addressed to be an extended  skyrmion-like mobile quasiparticle. However, at the same time it should'nt be forgotten that skyrmion corresponds to a collective state (excitation) of the whole system.

The boson addition or removal in the half-filled ($n=1$) boson system  can be a driving force for a nucleation of  a  multi-center ``charged'' skyrmions.  Such   {\it topological} structures, rather than  uniform phases predicted by the mean-field approximation, are believed to describe the evolution of the EBHM systems away from half-filling. It is worth noting that the multi-center skyrmions one considers as systems of skyrmion-like quasiparticles forming skyrmion liquids and skyrmion lattices, or crystals (see, e.g., Refs.\,\onlinecite{Timm,Green}).

\subsubsection{Dipole-quadrupole (pseudo)spin skyrmions}

In the continual limit for $J_1=J_2=J$ the Hamiltonian (\ref{ham7}) can be
transformed into the classical Hamiltonian of the fully $SU(3)$-symmetric
scale-invariant  model which can be rewritten as follows\,\cite{Nadya}:
$$
H_{isotr}=2J\int d^2{\bf r} \{
(\vec{\nabla}\Theta)^2+\sin^2\Theta(\vec{\nabla}\eta)^2+
$$
$$
\sin^2\Theta\cos^2\Theta\left[
	\cos^2\eta(\vec{\nabla}\Psi_1)^2+\sin^2\eta(\vec{\nabla}\Psi_2)^2\right]
$$
\begin{center}
\begin{equation}
	 	+\sin^4\Theta\cos^2\eta\sin^2\eta(\vec{\nabla}\Psi_1-\vec{\nabla}\Psi_2)^{2}\}\,
	, \label{iso}
\end{equation}
\end{center}
 where we have used the representation (\ref{fun}) and  introduced $\Psi_1=\Phi_1-\Phi_3,\Psi_2=\Phi_3-\Phi_2$. 
 The topological solutions for the Hamiltonian (\ref{iso}) can be classified  at least by
three topological quantum numbers (winding numbers): phases $\eta,\Psi_{1,2}$
can change by $2\pi$ after the passing around the center of the defect.
 The appropriate  modes may have very complicated topological structure due
 to the possibility for  one defect to have several different centers
 (while one of the phases $\eta,\Psi_{1,2,3}$ changes by $2\pi$ given
  one turnover
 around one center $(r_1,\varphi_1)$, other phases may pass around other
 centers $(r_i,\varphi_i)$). It should be noted that for such a center the winding numbers
  may take  half-integer  values. Thus we arrive at a large variety of topological structures to be solutions of the model.  Below we will briefly address two simplest classes of such
 solutions.
 One type of skyrmions  can be obtained given the trivial
phases $\Psi_{1,2}$. If these are constant, the ${\bf R}$ vector distribution
(see ($\ref{fun}$)) represents the skyrmion described by the usual formula
(\ref{sk}).
  All but one topological quantum numbers are zero for this class of solutions.
  It includes both dipole and quadrupole solutions: depending
  on selected constant phases one can obtain both "electric" and different
  "magnetic" skyrmions. The substitution $\Phi_1=\Phi_2=\Phi_3$ leads to
  the electric skyrmion which was obtained above as a solution of more general
   SU(3)-anisotropic model. Another
   example can be $\Phi_1=\Phi_2=0,\Phi_3=\pi/2$. This substitution
   implies  ${\bf b}\Vert Oz,{\bf a}\Vert Oxy,{\bf S}\Vert Oxy$, and
   ${\bf S}=\sin\Theta\cos\Theta \{\sin\eta,-\cos\eta,0\}$.
   Nominally, this is the in-plane spin vortex with a varying length of
   the spin vector
$$
|S|=\frac{2r\lambda|r^2-\lambda^2|}{(r^2+\lambda^2)^2}\,,
$$
which is zero at the circle $r=\lambda$, at the center $r=0$ and at
 the infinity $r\rightarrow \infty$,
and has maxima at  $r=\lambda(\sqrt 2 \pm 1)$. In addition to the non-zero
in-plane components of spin-dipole moment $\langle S_{x,y}\rangle$ this vortex is characterized by a non-zero
distribution of (pseudo)spin-quadrupole moments.
Here we would like to emphasize the difference between spin-1/2 systems in
 which there are such the solutions as in-plane vortices with the energy
  having a well-known logarithmic dependence on the size of the system
   and fixed spin length, and spin-1  systems in which the in-plane vortices
    also can exist but they may have a finite energy and a varying spin length.
     The distribution of quadrupole components associated with in-plane
      spin-1 vortex is non-trivial. Such solutions can be termed as
      "in-plane dipole-quadrupole skyrmions".

Other types of the simplest solutions with the phases
$\Psi_1=Q_1\varphi,\Psi_2=Q_2\varphi$ governed by two integer winding numbers
$Q_{1,2}$ and $\eta=\eta(r), \Theta=\Theta(r)$ are considered in Ref.\,\onlinecite{Nadya}.

\section{Conclusion}
Pseudospin formalism is shown to constitute a powerful method to study complex phenomena in interacting quantum systems.
We have focused here on the most prominent and intensively studied S=1 pseudospin formalism for extended bosonic Hubbard model with truncation of the on-site Hilbert space to the three lowest occupation states n = 0, 1, 2. The EHBM Hamiltonian is a paradigmatic model for the highly topical field of ultracold gases in optical lattices. At variance with standard EHBM Hamiltonian that seems to be insufficient to quantitatively describe the physics of bosonic systems the generalized non-Heisenberg effective pseudospin Hamiltonian, Eqs.(\ref{H})-(\ref{H2}) does provide a more deep link with boson system and  physically clear description of "the myriad of phases"\,  from uniform Mott insulating phases and density waves to two types of superfluids and supersolids. The  Hamiltonian could provide a novel starting point for analytical and computational studies of semi-hard core boson systems.
Furthermore, we argue that the 2D S=1 pseudospin system is prone to a topological phase separation and address different types of unconventional skyrmion-like structures, which, to the best of our knowledge, have not been analysed till now. The structures are characterized by a complicated  interplay of the insulating and the two superfluid phases with a single boson and boson dimers condensation, respectively.
Meanwhile we discussed the skyrmions to be  classical solutions of the continual isotropic models, however, this idealized object is believed to preserve their main features for strongly anisotropic (pseudo)spin lattice quantum systems. 
 Strictly speaking, the continuous model is relevant for discrete lattices only if we deal with long-wavelength inhomogeneities when their size is much bigger than the lattice spacing. In the discrete lattice the very notion of topological excitation seems to be inconsistent. At the same time, both quantum effects and the discreteness of the lattice itself do not prohibit from  considering the nanoscale (pseudo)spin textures whose topology and spin arrangement is that of a skyrmion\,\cite{skyrmion-cuprates,skyrmion}. 

We thank Yu. Panov and V. Konev for useful discussions. This research was supported in part  by the program 02.A03.21.0006 and  RFBR Grant No.  12-02-01039.

\end{document}